\documentclass[aip,reprint]{revtex4-2}
\usepackage{graphicx}
\usepackage{svg}
\usepackage{upgreek}
\usepackage{latexsym}
\usepackage{amsmath}
\usepackage{amssymb}
\usepackage{bm}
\usepackage{float}
\usepackage{color}

\renewcommand{\rho}{\varrho}
\renewcommand{\epsilon}{\varepsilon}
\renewcommand{\theta}{\vartheta}
\renewcommand{\phi}{\varphi}
\newcommand{\rb}{{\bf r}}

\newcommand{\kb}{{\bf k}}
\newcommand{\ub}{{\bf u}}

\usepackage{caption}
\usepackage{subcaption}
\usepackage{hyperref}
\hypersetup{colorlinks=true,linkcolor=blue,citecolor=blue,filecolor=blue,urlcolor=blue}
\usepackage[font=footnotesize, labelfont=bf]{caption}

\def\gtrless{\raise2.5pt\hbox{$>$}\llap{\lower2.5pt\hbox{$<$}}}
\def\gtrapprox{\raise2.5pt\hbox{$>$}\llap{\lower2.5pt\hbox{$\approx$}}}

\newcommand{\bsq}[1]{\begin{subequations}\label{#1}}
\newcommand{\esq}{\end{subequations}}
\newcommand{\beq}[1]{\begin{equation}\label{#1}}
\newcommand{\eeq}{\end{equation}}
\newcommand{\beqa}[1]{\begin{eqnarray}\label{#1}}
\newcommand{\eeqa}{\end{eqnarray}}

\draft 

\begin{document}


\title{Verification of Perrin’s theory of the motion of dilute spheroidal colloids} 



\author{John David Geiger}
\affiliation{Department Chemie, Universität Konstanz, 78457 Konstanz}
\author{Mohammed Alhissi}
\affiliation{Department Physik, Universität Konstanz, 78457 Konstanz}
\author{Markus Voggenreiter}
\affiliation{Department Chemie, Universität Konstanz, 78457 Konstanz}
\author{Matthias Fuchs}
\affiliation{Department Physik, Universität Konstanz, 78457 Konstanz}
\author{Andreas Zumbusch}
\email[]{andreas.zumbusch@uni-konstanz.de}
\affiliation{Department Chemie, Universität Konstanz, 78457 Konstanz}

\date{\today}

\begin{abstract}
Brownian motion is of central importance for understanding diffusive transport in biology, chemistry, and physics. For spherical particles, the theory was developed by Einstein, whereas a theoretical description of the motion of spheroids was given by F. Perrin. Here, we report the systematic verification of Perrin's theory 90 years after its publication. To this end, we synthesized oblate and prolate core-shell spheroids with different aspect ratios and tracked their three-dimensional diffusive motion in high dilution using confocal fluorescence microscopy. The experimental data for the dependence of translational and rotational diffusion on aspect ratio are in excellent agreement with the theoretical predictions. The crossover dynamics from anisotropic to isotropic diffusion as a hallmark for translation rotation coupling are also found as predicted. This verifies Perrin's theory as a cornerstone for understanding diffusive transport and underlines the excellent suitability of the particle system for testing more detailed theory. 
\end{abstract}


\maketitle 

At the beginning of the 19th century, Robert Brown published his observation of the random motion of pollen particles in water \cite{Brown1827}. The phenomenon he described and that is now known as as Brownian or diffusive motion, received little interest before the end of the century. In 1905 Einstein and shortly afterwards Smoluchowski and Langevin \cite{Einstein1905, Smoluchowski1906, Langevin1908} developed theoretical approaches explaining the observed motion of particles as a result of the thermal agitation of the solvent molecules. The quantitative measurements of J.B. Perrin that confirmed the theory were a cornerstone of the proof of the particle nature of matter\cite{Perrin1909}. For decades, numerous experiments in which the translational motion of spherical particles was directly monitored have left no doubt about the validity of the theoretical description and Brownian motion continues to play an important role for the understanding of numerous processes in biology, chemistry, and physics \cite{Frey2005}. 

To allow predictions about the diffusion behavior of molecules and particles of more complex shape, F. Perrin extended the theory originally developed for spherical particles to spheroids as a more general model\cite{Perrin1934, Perrin1936, Koenig1975}. Due to the anisotropic shape, their diffusion is not described by a single diffusion coefficient as for spherical particles, but by two diffusion tensors, one for translational diffusion and one for rotational diffusion. Moreover, the tensors are simple only in the particle-frame moving and rotating with  the colloid. Their determination required the solution of Stokes flow about ellipsoidal bodies \cite{Oberbeck,Edwardes,Jeffery}. Consequently, the theoretical description of this problem is significantly more intricate than that of the spherical case. Also experiments directly monitoring position and orientation of individual particles in solution proved to be much more challenging and only the development of new optical observation methods as well as of approaches to the synthesis of suitable spheroidal particles during the last two decades made experiments of this type possible. Using wide-field microscopy, Han et al. studied the translational and rotational diffusion of individual prolate spheroidal colloids in two dimensions \cite{Han2006, Zheng2010}. A special emphasis of this study was the experimental observation and theoretical description of the coupling between translational and rotational degrees of freedom caused by the anisotropic particle shape. Similar 2D experiments have since then been performed with colloidal dumbbells, clusters of spherical particles, and L-shaped particles \cite{Mayer2021,Anthony2008,Chakrabarty2013,Chakrabarty2014,Chakrabarty2016}. To capture details of translational and rotational coupling of colloidal dumbbells in 2D, the general angle- and position-dependent self intermediate scattering function was obtained theoretically \cite{Mayer2021}.

To follow the diffusion of individual prolate ellipoids with two different aspect ratios in a high viscosity solvent in three dimensions, Mukhija and Solomon used confocal fluorescence microscopy \cite{Mukhija2007}. The same approach was used to image the diffusion of clusters of spherical particles \cite{Hunter2011,Kraft2013}, whereas holographic microscopy served to monitor the 3D diffusion of colloidal rods and cylinders \cite{Wang2014}, and depolarized dynamic light scattering gave information on colloids with partially crystalline internal structure \cite{Degiorgio1994}. Since these 3D measurements were limited to determining the diffusion coefficients in the laboratory frame, no coupling between translation and rotation was observed. In 3D, the observation of the crossover from anisotropic to isotropic diffusion in the particle-frame as a hallmark of this coupling was reported only recently for the diffusion of Au nanorods that was monitored using darkfield microscopy \cite{Molaei2018}.

\begin{figure*}[ht]
\includegraphics[width=6.3in]{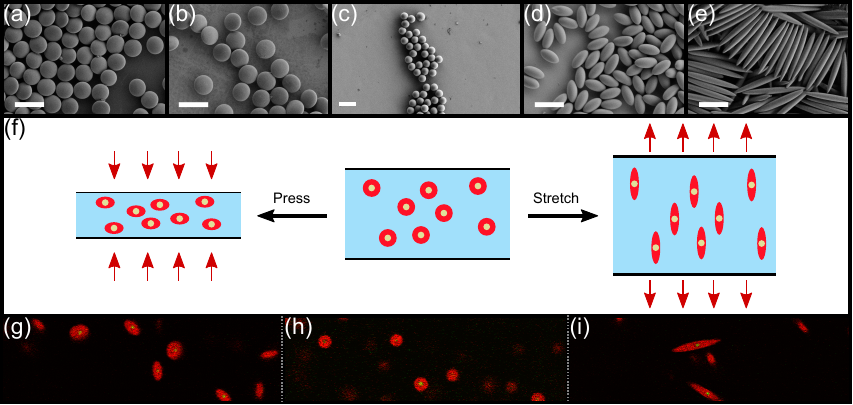}
\caption{\label{fig1}Particle system. (a-c) Scanning electron microscopy (SEM) images of oblate ellipsoids with aspect ratios b/a=2.1 (a) and b/a=2.4 (b), spherical precursor particles (c) and prolate ellipsoids with aspect ratios a/b=2.0 (d) and a/b=9.4 (e). Scale bars: 10~$\mu$m. (f) Schematic representation of the thermomechanical pressing (left side) and stretching (right side) procedure to produce oblate and prolate spheroids. Confocal fluorescence images of oblate spheroids with aspect ratio $b/a=2.1$ (g), spherical precursors (h) and prolate spheroids with the aspect ratio $a/b=6.6$ (i). Core signals in green, shell signals in red.}
\end{figure*}

Despite its great importance for understanding transport phenomena in general, an experimental verification of F. Perrin's theory for the diffusion of spheroids is still missing. The main aspects of such a confirmation are experimental data on the dependence of translational and rotational diffusion on the particles' aspect ratios and the experimental observation of translation-rotation coupling. In 2D, a few experimental data on different aspect ratios are available \cite{Han2006}, whereas in 3D only results of one study with measurements of two aspect ratios is available. Data on oblate spheroids so far are missing altogether. Here, we report results of confocal fluorescence microscopy experiments of the diffusion of spheroidal colloids diffusing in 3D. Apart from measurements of spherical particles, our data comprise diffusion measurements of particles with eight different aspect ratios, four of oblate and four of prolate shape. The temporal resolution of our measurements is high enough to allow for monitoring the crossover from anisotropic to isotropic translational diffusion. For the analysis of the latter, we recapitulate the most important aspects of the recent theoretical description of 2D diffusion \cite{Mayer2021} for 3D and perform Brownian dynamics calculations.

Advances in controlled particle synthesis gives access to a variety of well-defined particle geometries like rods\cite{Hagemans2019,Fernandez2018}, dumbells\cite{Hueckel2018}, bananas\cite{Fernandez2021}, discs\cite{Sacanna2013}, cubes\cite{Meijer2017}. Based on a method developed by Ho et al., also spheroids with prolate and oblate shape can be prepared by controlled stretching or squeezing of spherical precursor particles\cite{Ho1993,Voggenreiter2020}. The colloidal particles used in our study possess a core-/shell-geometry with crosslinked poly(methyl methacrylate) (PMMA) cores and non-crosslinked PMMA shells. The particle synthesis is based on a method first published by Antl \emph{et al.}\cite{Antl1986}. We modified this method to obtain particles in which both, cores and shells, are labelled with different fluorescent dyes. This allows the simultaneous detection of the position and the orientation of spheroidal particles in two different detection channels. We obtained spherical PMMA-core/PMMA-shell particles with a diameter of 4.69~$\mu m$ with a polydispersity of 2.3~\%. The particles were sterically stabilized by covalent binding of the graft copolymer polyhydroxystearic acid-$\emph{g}$-PMMA \cite{Hollingsworth2006} to achieve hard potentials. To ensure that the particles investigated possessed the same volume, all spheroidal particles with different aspect ratios were produced from this batch of particles using a thermomechanical method (Fig.\ref{fig1}). Prolate spheroidal particles were obtained by stretching \cite{Keville1991, Zhang2011}, whereas oblate particles were synthesized by pressing \cite{Voggenreiter2020}.

For the diffusion measurements, the particles were suspended in a mixture of cyclohexylbromide (CHB) and \emph{cis}-decalin to achieve density and refractive index matching of particles and solvent\cite{Yethiraj2003}. Tetrabutylammonium bromide (TBAB) was added to screen charges and to minimize electrostatic interactions between particles. The amount of solvent was chosen as such that volume fractions $\upphi$ were below 2\,\%. Fluorescence images of the diffusing particles were acquired using a commercial confocal microscope (TCS SP5 Leica Microsystems). For the diffusion measurements, 3D image stacks were recorded with lag times of 4~s. From the imaging data, spatio-temporal trajectories of individual particles were obtained and analysed using a home-written software\cite{Roller2018}. While the detection of the core signal served to determine the particles' positions, their orientation was derived from the shell signals. 

The trajectories serve as the basis for the analysis of the particles' translational and rotational diffusion. Fitting the mean squared displacement (MSD) $\langle\Delta r^2(t)\rangle$ to $\langle\Delta r^2(t)\rangle=6D_tt+6\epsilon_t^2$ and the mean squared angular displacement (MSAD) $\langle\Delta \theta^2(t)\rangle$ to $\langle\Delta \theta^2(t)\rangle=4D_{r}t+2\epsilon_r^2$, we obtain the translational diffusion coefficients $D_t$ of spheres, prolate and oblate ellipsoids and the rotational diffusion coefficients $D_r$ of prolate and oblate ellipsoids. The curve fittings also yield the translational and rotational measurement uncertainties $\epsilon_t$ and $\epsilon_r$ that are reported for each sample in the supplemental material \cite{Besseling2015}.

For the low particle concentrations used in the experiments, the translational diffusive motion of the particles is expected to follow $D = \mu k_B T$ with the particles' mobility $\mu$, Boltzmann's constant $k_B$, and temperature T\cite{Einstein1905,Smoluchowski1906,Sutherland1905}. For low Reynolds numbers, the drag coefficient $\gamma_t = 1/\mu$ and $D_t=\frac{k_BT}{\gamma_t}$. For spherical particles with a hydrodynamic radius $r$ in viscous fluids with the dynamic viscosity $\eta$, $\gamma_t=6\pi\eta r$ which yields the well known Stokes-Einstein equation. In contrast to spherical particles, where friction and diffusion coefficients are isotropic scalars, the diffusion tensor of spheroidal particles in the particle-frame contains a coefficient 
\begin{equation}\label{eq1}
D_{t,\parallel}=\frac{k_BT(2-p^2)g(p)-1}{8\pi\eta a(1-p^2)}\,,
\end{equation}
for motion parallel to their major semi-axis \cite{Kim2005,Hoffmann2009}, whereas for motions perpendicular to their major semi-axis the coefficient reads
\begin{equation}\label{eq2}
D_{t,\perp}=\frac{k_BT(2-3p^2)g(p)+1}{16\pi\eta a(1-p^2)}\,.
\end{equation}
In the last expressions, 
\begin{equation}\label{eq3}
g(p)=\frac{\log{\frac{1+\sqrt{1-p^2}}{p}}}{\sqrt{1-p^2}}
\end{equation}
for $p<1$ and
\begin{equation}\label{eq4}
g(p)=\frac{\arctan{\sqrt{p^2-1}}}{\sqrt{p^2-1}}
\end{equation}
for $p>1$. Here, $p=\frac{b}{a}$ is the aspect ratio defined as the length ratio of the minor semi-axis $b$ and the major semi-axis $a$. From these expressions, the translational diffusion coefficient for spheroids in the laboratory frame is calculated to be
\begin{equation}\label{eq5}
D_t=\frac{D_{t,\parallel}+2D_{t,\perp}}{3} \,.
\end{equation}
The diffusion coefficient for rotations of rigid spheroids about their minor semi-axis $b$ as derived by Perrin \cite{Perrin1934,Koenig1975} is given by \cite{Kim2005,Hoffmann2009}
\begin{equation}\label{eq6}
    D_{r}=\frac{3k_BT(2-p^2)g(p)-1}{16\pi\eta a^3(1-p^4)}\,.
\end{equation}

\begin{figure}[ht]
\includegraphics[width=3.3in]{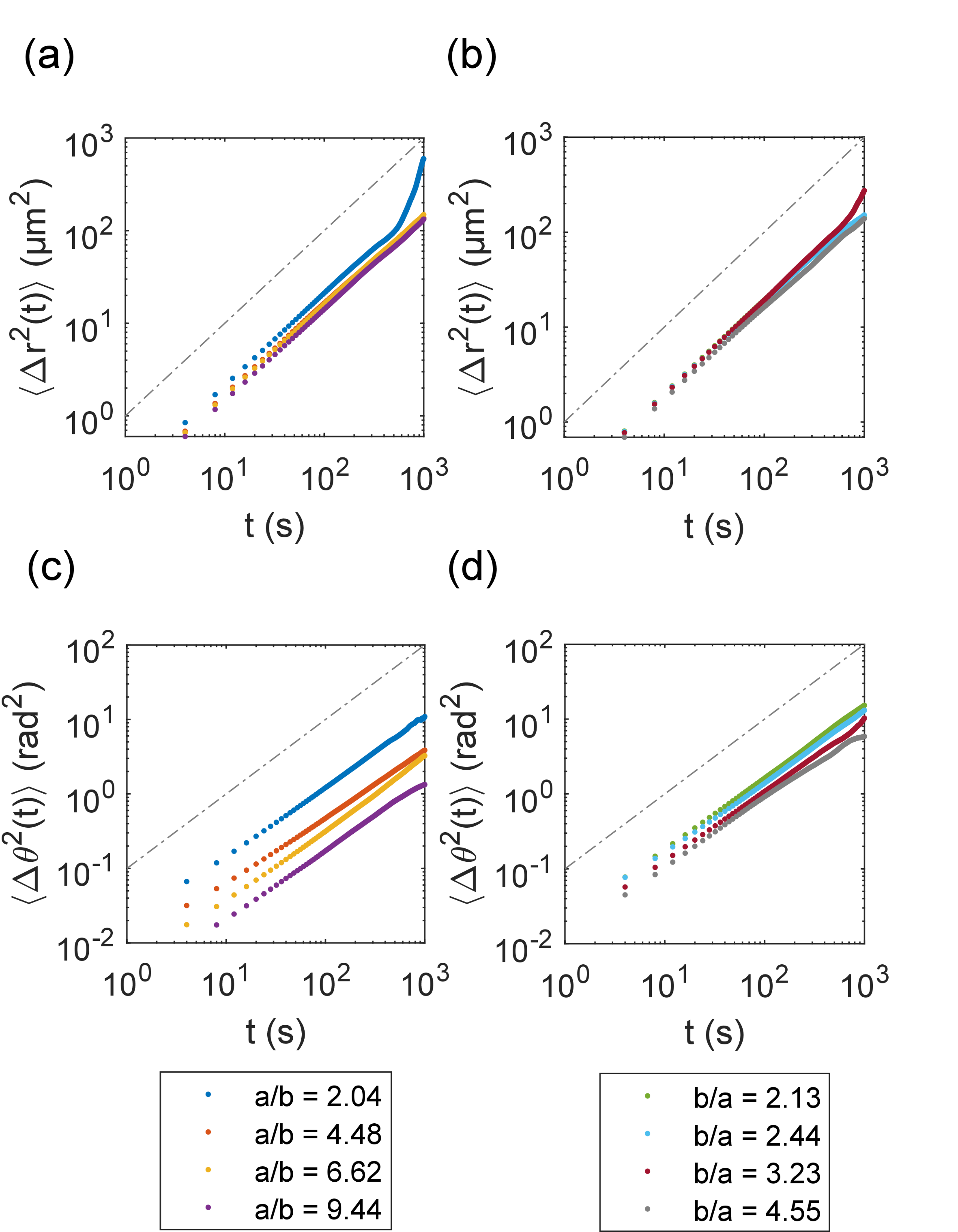}
\caption{\label{fig2}Mean squared displacement curves of prolate (a) and oblate (b) spheroids. Mean squared angular displacement curves of prolate (c) and oblate (d) spheroids. The dot-dashed with slope 1 line serves as a guide to the eye.}%
\end{figure}

On basis of the recorded trajectories, we first analyzed MSDs and MSADs of prolate and oblated spheroids with different aspect ratios (Fig.~\ref{fig2}). Linear dependence on time, as expected for free diffusion, is found in all cases. Since the friction coefficients increase with aspect ratio, theory predicts the observed decrease of both the translation and the rotation diffusion coefficients with aspect ratio. By fitting the MSD and MSAD data, we obtained translational and rotational diffusion coefficients. The aspect ratio dependence of both, the translational and rotational diffusion coefficients $D_t$ and $D_r$ thus determined, is in excellent agreement with the F. Perrin's theory as given by equations~\ref{eq1}-\ref{eq6} (Fig.~\ref{fig4}~a).
One should note that the determination of the rotational diffusion coefficients is less error prone than that of their translational counterparts, since the latter suffer more strongly from residual drift in the sample chamber. 

\begin{figure*}[htb]
\includegraphics[width=6.3in]{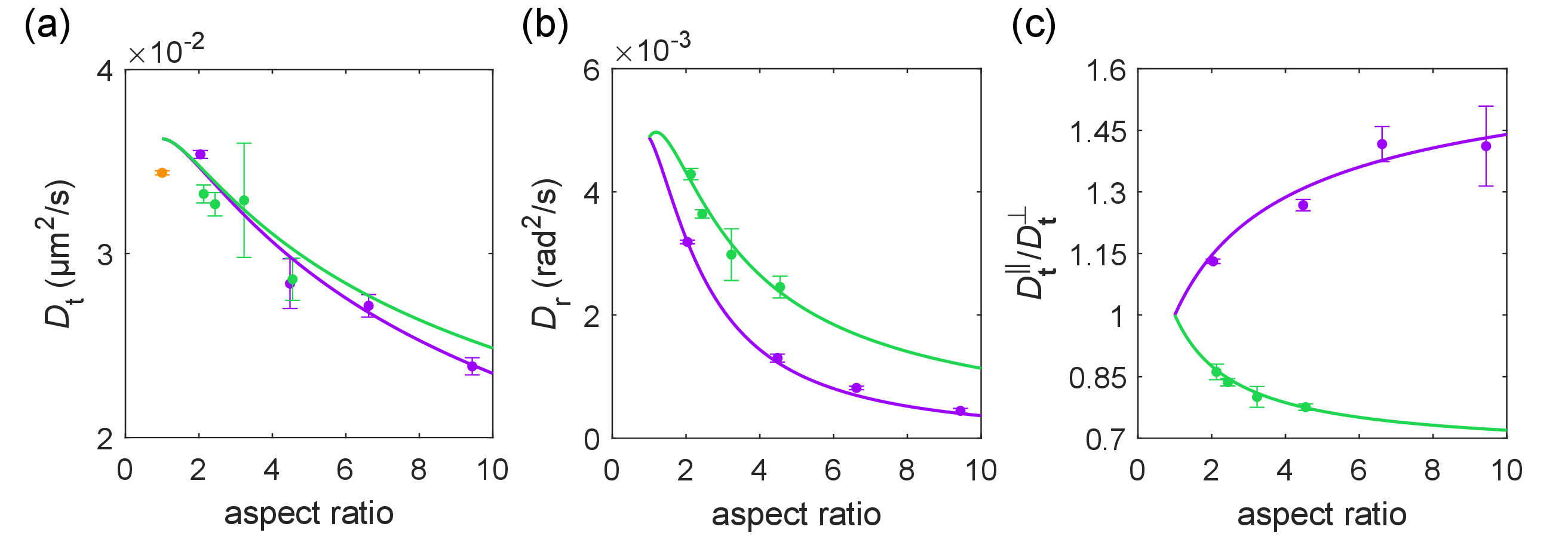}
\caption{\label{fig3} Comparison between experimentally determined diffusion coefficients and theoretical predictions for different spheroidal particle aspect ratios. (a) Experimentally determined translational diffusion coefficients $D_t$ for prolate (purple, aspect ratio $a/b$) and for oblate (green, aspect ratio $b/a$) spheroids. Solid lines: theoretical prediction using equations \ref{eq1}-\ref{eq5}. The equations were fitted to the experimentally determined translational diffusion coefficients of both prolate and oblate ellipsoids using the length of the major semi-axis as free parameter. (b) Experimentally determined rotational diffusion coefficients D$\bf{_r}$ for prolate (purple, aspect ratio $a/b$) and oblate spheroids (green, aspect ratio $b/a$). Solid lines: theoretical predictions using equations \ref{eq1}-\ref{eq4} and \ref{eq6}. (c) Experimentally determined ratio of $D_{t, \parallel}/D_{t, \perp}$ for prolate (purple) and oblate (green) spheroids. Solid lines: Theoretical predictions using equations \ref{eq1}-\ref{eq4}.}
\end{figure*}

\begin{figure}[ht]
\includegraphics[width=3.0in]{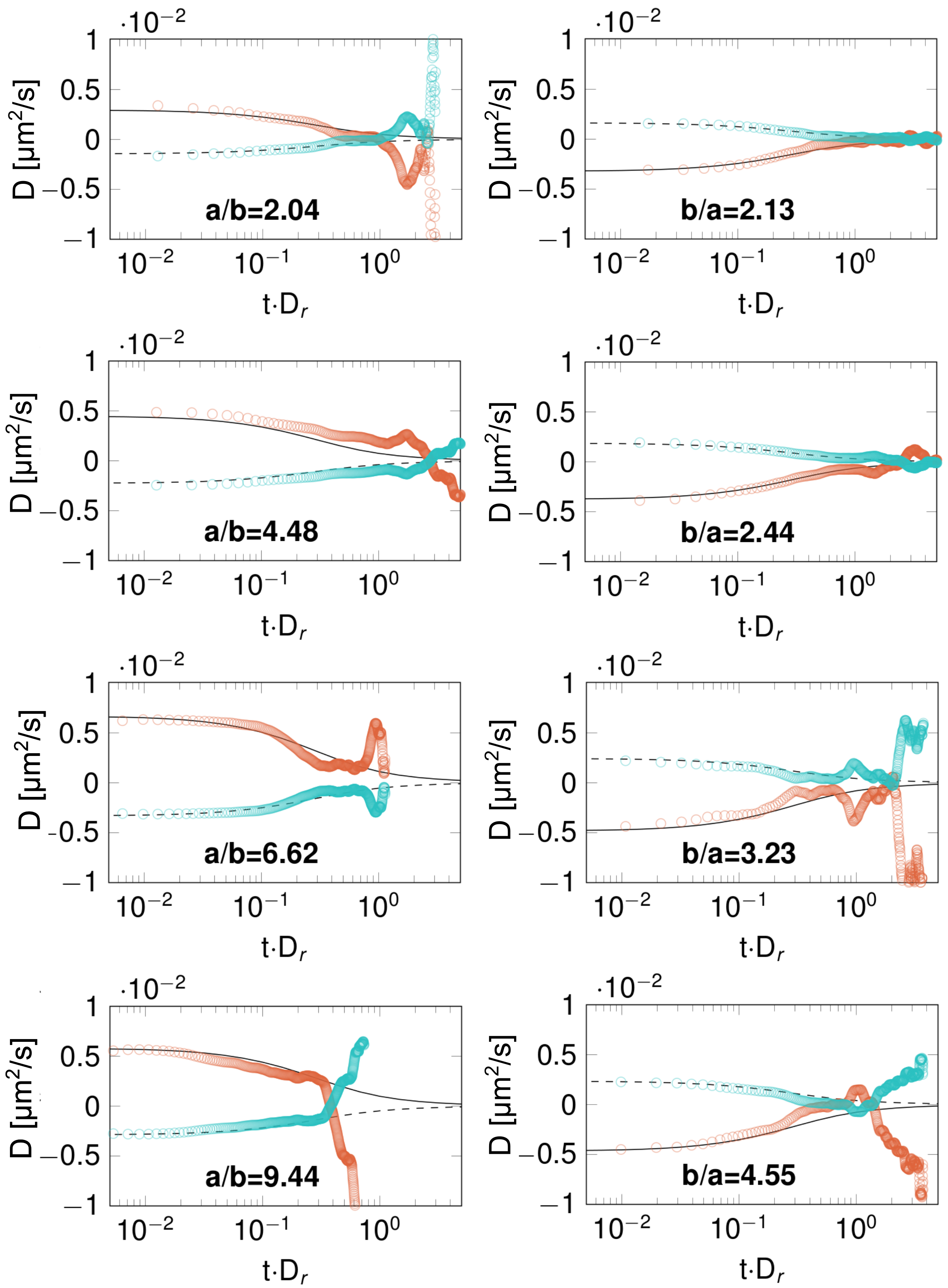}
\caption{\label{fig4}Crossover of anisotropic to isotropic diffusion for prolate (left column) and oblate (right column) spheroids. Red circles: $D_{\parallel,\theta_0} - \langle \Delta r^2\rangle/6t$; blue circles: $D_{\perp,\theta_0} - \langle \Delta r^2\rangle/6t$. The $\langle \Delta r^2\rangle/6t$ term was subtracted to account for residual drift. Solid lines are fits to eq.~\ref{eq7}, whereas dashed lines are fits to eq.~\ref{eq8}.}
\end{figure}

In addition to the analysis of the particle motion in the laboratory frame, imaging of individual particle trajectories allows for the determination of diffusion coefficients in the body frame. To this end, the translational motion of every particle is divided up into a translational displacement $r_\parallel$ parallel to its major semi-axis $a$ and a translational displacement $r_{\perp}$ perpendicular to its minor semi-axes $b$. Here, the initial orientation of each ellipsoidal particle is used as reference system. One thus obtains $\langle\Delta {r^2}_\parallel(t)\rangle$ and $\langle\Delta {r^2}_\perp(t)\rangle$ from which the respective diffusion coefficients $D_{t,\parallel}$ and $D_{t, \perp}$ were derived by linear regression (Fig.~\ref{fig4}~c). Plotting the dependence of $D_{t, \parallel}/D_{t,\perp}$ on aspect ratio, also in this case we find excellent agreement between theory and experiment for oblate as well as for prolate spheroids (see Fig.\,\ref{fig4}\,c). 
An important effect of the anisotropic diffusion of spheroidal particles is the coupling between translational and rotational diffusion. This can be tested by examining the short time behavior of $D_{t,\parallel}$ and $D_{t, \perp}$. While initially, $D_{t,\parallel} > D_{t, \perp}$, rotational motion leads to a crossover from anisotropic to isotropic diffusion approaching $D_t$ at long times. F. Perrin found
\begin{align}
\langle r^2_{\parallel, \theta_0} (t)\rangle & = 2 D_t t + \frac 29 \frac{\Delta D}{D_r} (1-  e^{-6 D_r t} ) \label{eq7}\\
\langle r^2_{\perp, \theta_0} (t)\rangle & = 2 D_t t - \frac 19 \frac{\Delta D}{D_r} (1-  e^{-6 D_r t} ) \label{eq8}
\end{align}
with $\Delta D = D_\parallel - D_\perp$ and the initial angle $\theta_0$ aligned with the major semi-axis\cite{Perrin1936}. We confirmed these expressions using the methodology developed by Mayer et al. for 2D systems (supplemental material). This crossover has been reported previously for 2D diffusion of prolate spheroids \cite{Han2006} and of dumbbells \cite{Mayer2021}, as well as for Au nanorods with one aspect ratio in 3D\cite{Molaei2018}. Fig.~\ref{fig4} shows the crossover for oblate and prolate spheroids with two different aspect ratios. Again the experimental data coincide very well with theory. Translation-rotation coupling is also expected to show up at intermediate times on the crossover scale as a non-Gaussian distribution of the laboratory-frame displacements parallel and perpendicular to an initial orientation vector of the spheroids. This has been reported previously in 2D, where the strong confinement resulted in $D_{t, \parallel}/D_{t, \perp} \approx 4$. In our case, no signs for non-Gaussian displacement distributions were found. This can be explained by the fact that for 3D systems, the maximum value for $D_{t, \parallel}/D_{t, \perp} = 2$ for $a/b \to \infty$. For our particles with the highest aspect ratio but still reasonably short rotational relaxation times, we calculate $D_{t, \parallel}/D_{t, \perp} = 1.27$ ($a/b = 4.48$). This ratio probably is too small to lead to detectable deviations from Gaussian distributions. We also verified by Brownian dynamics simulations that finite density corrections can be neglected in our experiments (see supplemental material).

In summary, we report measurements confirming F. Perrin's theory describing the Brownian motion of spheroidal particles that is of extraordinary importance for the description of transport phenomena. After synthesizing colloidal particles with identical volumes and eight different aspect ratios we could use confocal fluorescence microscopy to obtain trajectories of individual particles. This allowed us to derive diffusion coefficients in the laboratory and in the particles' body frame. These and data obtained on the dynamics of the crossover from anisotropic to isotropic diffusion are in excellent agreement with theory. This verifies Perrin's theoretical work 90 years after its publication and shows that the theory correctly captures also subtle effects like the coupling of translation and rotation. Our work also shows that the particle system employed is an attractive model for investigating complex 3D transport in fluids. It promises to be well suited for testing more detailed theory that is currently developed \cite{Mayer2021}.\\

\section*{Acknowledgements}

This work was funded by the German Reearch Council as project C07 in SFB1432.

\section*{References}
%

\newpage

\textbf{\Large{Supplemental Material}}

\section{Experimental Section}

\subsection{Synthesis of core-/shell-particles}
The colloidal particles used possessed a core-/shell-geometry with crosslinked poly(methyl methacrylate) (PMMA) cores and non-crosslinked PMMA shells. Our particle synthesis is based on a method first published by Antl et al. \cite{Antl1986}.  We modified this method to obtain particles in which both, cores and shells, are labelled with a fluorescent dye. In brief, first core particles were synthesized via dispersion polymerisation. The dye, a vinyl-functionalized derivative of Quasar 670 (LGC Biosearch Technologies, USA), was copolymerized during this reaction and the cores are crosslinked using ethylglycoldimethylacrylate. By using the core particles as seeds in a seeded dispersion polymerisation, a PMMA shell layer was grown onto the particles. In this step, the Bodipy dye  (4,4-difluoro-8-(4-methacrylatephenyl)-3,5-bis-(4-methoxyphenyl)-4-bora-3a,4a-diaza-s-indacene), was copolymerized to fluorescently label the shell of the particles. The dye itself was synthesized according to literature.\cite{Baruah2005} The resulting particles tended to have a rough surface structure. They were smoothed by dispersing and stirring them in a solvent mixture of decalin and acetone for 1 hour \cite{Klein2015} which lead to spherical PMMA core/PMMA shell particles with smooth surfaces.

\subsection{Thermomechanical stretching procedure}
The spherical PMMA-core/PMMA-shell particles were then deformed by thermomechanical methods to yield prolate or oblate ellipsoids. To get prolate particles, we used a stretching procedure that has first been described by Keville and coworkers\cite{Keville1991}. Initially, the particles were suspended in a mixture of polydimethylsiloxane (PDMS), hexane, the crosslinker (poly(dimethyl-siloxane-co-methylhydrosiloxane)), and tin(II)2-ethylhexanoate as a catalyst. This suspension was cast into a metal frame. After evaporation of hexane, the mixture was heated to 120$^{\circ}$~C for 6 hours. The resulting crosslinked PDMS film was cut into stripes that are clamped on both sides before being heated to 165$^{\circ}$~C. To obtain particles with different aspect ratios, the stripes were stretched with different stretching factors. After cooling down, the PDMS matrix was degraded using a solution of sodium methanolate and isopropanol in hexane. The released particles were then washed with decaline in several steps. Finally, the graft copolymer PHSA-$\emph{g}$-PMMA, that was synthesized according to literature \cite{Hollingsworth2006}, was covalently attached to the particle surfaces.

\subsection{Thermomechanical pressing procedure}
Oblate ellipsoids were generated by pressing the spherical core-shell particles. This method has recently been described in detail \cite{Voggenreiter2020}. At the beginning, the spherical core-shell-particles were embedded in a PDMS matrix as described in the previous section. Thereafter, the particle film was placed between two metal plates and pressed with different pressing factors to obtain prolate ellipsoidal particles with the desired aspect ratios. The subsequent steps of degrading, washing, and covalent attachment of the stabilizer on the particles surface were the same as those described in the previous section.

\subsection{Particle characterization}

\begin{table}[h]
\caption{Polydispersity index (PDI) for the sahpes of the synthesized particles. The values were derived from scanning electron microscopy data.}
\begin{tabular}{l|cc}
aspect ratio & spheroid type & polydispersity $a/b$ [\%]\\
\hline
$a/b = 2.04$ & prolate & 5.6\\
$a/b = 4.48$ & prolate & 6.5\\
$a/b = 6.62$ & prolate & 6.8\\
$a/b = 9.44$ & prolate & 6.5\\
$b/a = 2.13$ & oblate & 10.2\\
$b/a = 2.44$ & oblate & 10.8\\
$b/a = 3.23$ & oblate & 12.5\\
$b/a = 4.55$ & oblate & 14.9
\end{tabular}
\end{table}
\subsection{Sample preparation}

For the diffusion measurements, the particles were suspended in a mixture of 85wt-\% cyclohexylbromide (CHB) and 15wt-\% of \textit{cis}-decalin to achieve density and refractive index matching of particles and solvent mixture \cite{Yethiraj2003}. Tetrabutylammonium bromide (TBAB) was added to the mixture until saturation to screen charges and to minimize electrostatic interactions between particles. Since PMMA particles are known to swell in CHB, the particles were left in the mixture for 14 days to reach saturation. Afterwards the particles were density matched at the measurement temperature of 23$^{\circ}$~C by centrifugation and the adjustment of the amount of decalin or CHB, respectively. The amount of solvent was chosen such that volume fractions were below 2~\%. The glass chambers had dimensions of 75x25x3\,mm$^3$ with a cylindrical cavity (diameter 2.5\,mm, height 2.8\,mm) on one side and a cylindrical cavity (diameter 0.5\,mm, height 8\,mm) on the other side. Both were connected in the middle of the chamber. The sample chambers were sealed with glass coverslides (18x18x0.17\,mm$^3$, Marienfeld) at least 12 hours before measurement using two-component epoxy adhesive (UHU Plus Sofortfest).

\subsection{Confocal imaging}
Image were acquired with a confocal fluorescence laser scanning microscope (TCS SP5, Leica Microsystems) with a resonant scanner (8000\,Hz, bidirectional scanning mode) and a glycerol immersion objective (63x magnification, 1.3 NA). The microscope objective was covered by a box which connected to a temperature stabilization system (Ludin Cube 2, Life Imaging Services) which held the temperature at 23\,$\pm$\,0.05$^{\circ}$~C to ensure density matched suspensions. A helium-neon laser ($\uplambda$ = 633\,nm) and an argon laser ($\uplambda$ = 514\,nm) were used to excite the fluorophores in the particle cores and shells and the fluorescent light was simultaneously detected in two separate detection channels. For the diffusion measurements, 3D image stacks (1024x256x100 voxels) with pixel sizes of dx=dy=141.3\,nm and dz=210\,nm resulting in 3D image volumes of 144.7x36.2x21.0$\upmu$m$^3$ were recorded with lag times of t=4~s. The image stacks were recorded with distances of at least 25\,$\upmu$m to the measurement chamber walls. To allow equilibration, all samples are put onto the objective 16 hours before the measurement starts.

\subsection{Data processing}
\subsubsection{Tracking}
For particle detection and tracking, we used an algorithm that takes advantage of the particles' core-shell structure \cite{Roller2018}. The software and all other data processing scripts used for the calculation of diffusion parameters is written Matlab R2019a (The MathWorks, Inc.). The algorithm uses fluorescence imaging data of the cores to derive the 3D positions. The anisotropic shape of the shells, by contrast, serves for the detection of the 3D orientation of the prolate and oblate ellipsoidal particles. From the positions and orientations, the algorithm generates the respective single-particle trajectories. 

\subsubsection{Detection accuracies}
The trajectories served as the basis for the analysis of the particles' translational and rotational diffusion. Fitting the mean squared displacement (MSD) $\langle\Delta r^2(t)\rangle$ and the mean squared angular displacement (MSAD) $\langle\Delta \theta^2(t)\rangle$ with equations \eqref{smeq01} and \eqref{smeq02}, respectively, we obtained the translational diffusion coefficients $D_t$ of spheres, prolate and oblate ellipsoids and the rotational diffusion coefficients $D_r$ of prolate and oblate ellipsoids. 

\begin{equation}\label{smeq01}
    \langle\Delta r^2(t)\rangle=6D_tt+6\epsilon_t^2
\end{equation}
\begin{equation}\label{smeq02}
    \langle\Delta \theta^2(t)\rangle=4D_{r}t+2\epsilon_r^2
\end{equation}

Here, $\epsilon_t$ and $\epsilon_r$ are the respective position and orientation determination accuracies \cite{Besseling2015}. For the individual aspect ratios, we found the following values:\\

\begin{table}[h]
\caption{Detection accuracies $\epsilon$ for the different particles.}
\begin{tabular}{l|cc}
aspect ratio & $\epsilon_t$ [nm] & $\epsilon_r$ [degrees]\\
\hline
$a/b = 2.04$ & 13 & 5.0\\
$a/b = 4.48$ & 39 & 4.1\\
$a/b = 6.62$ & 37 & 2.7\\
$a/b = 9.44$ & 67 & 2.1\\
$b/a = 2.13$ & 32 & 3.8\\
$b/a = 2.44$ & 40 & 5.4\\
$b/a = 3.23$ & 20 & 3.8\\
$b/a = 4.55$ & 37 & 3.0
\end{tabular}
\end{table}

\section{Perrin equation}

For reference, Perrin's equations are derived for the conditioned anisotropic mean squared displacements (MSD) for various directions relative to the initial orientation of the ellipsoidal colloid. This follows work in two dimension \cite{Mayer2021} with some techniques from three \cite{Kurzthaler2016} .

For an isotropic diffusor \cite{Berne}, the pdf of displacements $\rb$ and orientations $\ub$ (with $\ub$ a unit vector pointing along the main axis of the particle) shall be denoted as $P(\rb,\ub,t|\theta_0)$. It is conditioned that the initial orientation has the angle $\theta_0$ to the laboratory $\bf\hat z$-axis, thus: 
\beq{smeq1}
P(\rb,\ub,t=0|\theta_0) = \delta^3(\rb)\; \delta( \cos{\theta} - \cos{\theta_0} ) \; \frac{1}{2\pi}\;. 
\eeq
Any initial azimuthal angle $\phi_0$ is equally probable.
The Smoluchowski-Perrin equation gives the time evolution of the pdf:
\beq{smeq2}
\partial_t \; P(\rb,\ub,t|\theta_0) = \left[  D_r \Delta_\ub + D_\perp \partial_\rb^2 + \Delta D (\partial_\rb\cdot \ub)^2 \right]\; P \;,
\eeq
where $\Delta_\ub= \partial_\eta (1-\eta^2)\partial_\eta + \frac{1}{1-\eta^2} \partial_\phi^2$ is the angular part of the Laplacian, with abbreviation $\eta=\cos{\theta}$. $D_r$ is the rotational diffusion coefficient, $D_\perp$ the translational  one perpendicular to the main axis, and $\Delta D=D_\|-D_\perp$ the anisotropy in the translational diffusivities. Later the mean translational diffusivity $\bar{D}=\frac13(2D_\perp+D_\|)$ will appear. Note that the setup corresponds to a translational diffusion tensor of the form ${\bf D}=D_\|\ub\ub+D_\perp({\bf 1}-\ub\ub)$.

For the MSD, one requires the Legendre polynomials $P_n(\eta)$, which obey the ordinary differential equation (ODE)
\beq{smeq3}
\Delta_\ub \; P_n(\eta) = - n (n+1)\; P_n(\eta)\;, 
\eeq
and form a complete set in one dimension according to
\beq{smeq4}
\delta( \eta - \eta_0 ) =  \sum_{n=0}  \frac{2n+1}{2}\; P_n(\eta) \, P_n(\eta_0) \;.
\eeq
Their recursion relation will be useful, which is valid for $n\ge1$:
\beq{smeq5}
\eta \; P_n(\eta) = \frac{n+1}{2n+1} P_{n+1}(\eta) + \frac{n}{2n+1} P_{n-1}(\eta) \;.
\eeq 
Actually, we will need only the simple consequence $\eta^2P_0(\eta) = \frac23 P_2(\eta) + \frac 13 P_0(\eta)$ obtained by iteration. 

The generalized self intermediate scattering functions $F_n(\kb,t|\theta_0)$  shall be defined as they encode the rotational and translational motion:
\beq{smeq6}
F_n(\kb,t|\theta_0) = \int d^2\ub\, P_n(\eta) \int d^3\rb \, e^{-i\kb\cdot\rb}\; 
P(\rb,\ub,t|\theta_0) \;,
\eeq
where $\int d^2\ub=\int_0^{2\pi}d\phi\int_{-1}^1d\eta$.
Their initial values are $F_n(\kb,t=0|\theta_0) = P_n(\eta_0)$. 
The angle averaged MSD follow from (using $P_0(\eta)=1$):
\beq{smeq7}
(i \frac{\partial}{\partial k_i})^2 \; 
F_0(\kb,t|\theta_0)\big|_{\kb=0} = \delta r_i^2(t)\big|_{\theta_0}\; .
\eeq
They depend on the polar angle $\theta_0$ of the initial orientation $\ub_0\cdot\hat{\bf z}=\cos{\theta_0}=\eta_0$ in lab frame. 

Performing the integrations given in Eq.~\eqref{smeq6}, the Perrin equation can be transformed into coupled ODE for the scattering functions. 
Setting $\tilde{P}(\kb,\ub,t|\theta_0)= \int d^3\rb \, e^{-i\kb\cdot\rb} P(\rb,\ub,t|\theta_0)$ the scattering functions $F_n(\kb,t|\theta_0)$ obey:
\beqa{smeq8}
\partial_t F_n(\kb,t|\theta_0)=\int d^2\ub\, P_n(\eta)   \left[  D_r \Delta_\ub + D_\perp k^2 + \Delta D (\kb\cdot \ub)^2 \right] 
\tilde{P}\nonumber\\
= - \int d^2\ub\, P_n(\eta)   \left[  D_r n(n+1) + k^2(D_\perp + \Delta D \eta^2) \right] 
\tilde{P}\;,\quad
\eeqa
where the wavevector lies along the lab-z-axis, $\kb=k\hat{\bf z}$.
Now, the recursion relation \eqref{eq5} is required to derive by iteration:
\beqa{smeq9}
\eta^2 \; P_n(\eta)  &=& \alpha^+_n   P_{n+2}(\eta) +\alpha^-_n   P_{n-2}(\eta) + \alpha_n   P_{n}(\eta) \\\nonumber 
&=& 
\frac{(n+1)(n+2)}{(2n+1)(2n+3)} P_{n+2} + \frac{n(n-1)}{(2n+1)(2n-1)} P_{n-2 }  
\\\nonumber 
&& + \left( \frac{(n+1)^2}{(2n+1)(2n+3)} + \frac{n^2}{(2n+1)(2n-1)} \right)\; P_{n } \;.
\eeqa
Setting this into Eq.~\eqref{smeq8} gives finally:
\beqa{smeq10}
[ \partial_t + n(n+1)D_r]  F_n(\kb=k\hat{\bf z},t|\theta_0) = \qquad\qquad\\\nonumber - k^2 \big\{   [ D_\perp + \Delta D \alpha_n ] F_n  + \Delta D (\alpha^+_n F_{n+2} +  \alpha^-_n F_{n-2} ]\big\}\,.\quad
\eeqa

Because of Eq.~\eqref{smeq7}, the scattering functions are required only up to quadratic order in wavevector. Clearly, from Eq.~\eqref{smeq10} and the initial value follows the decay of the scattering function at $\kb=0$:
\beq{smeq11}
F_n(\kb={\bf 0},t|\theta_0) = P_n(\eta_0)\; e^{-n(n+1) D_r t}\; .
\eeq
Continuing with the MSD, again because of  Eq.~\eqref{eq7}, only $F_{n=0}$ is required, as $F_0(\kb=k\hat{\bf z},t|\theta_0)-F_0({\bf 0},t|\theta_0) = -\frac{k^2}{2}  \delta r_i^2(t)\big|_{\theta_0} +\ldots$. From Eqs.~\eqref{smeq10} and \eqref{smeq11} and with $\alpha^+_0=\frac 23$, $\alpha_0=\frac 13$, and $\alpha^-_0=0$ (from Eq.~\ref{smeq9}) follows:
\beq{smeq12}
\partial_t \delta r_i^2(t)\big|_{\theta_0}  = 2 \bar{D} + 2 \Delta D (\eta_0^2-\frac 13 ) e^{-6 D_r t}\;.
\eeq
The relaxation of the initial orientation encoded in $\eta^2_0=\cos{\theta_0}^2$  needs to get forgotten by rotational diffusion, before isotropic diffusion can set in with the average diffusion coefficient $\bar D$. Note the different exponential decay to 2D, where
$e^{-n^2D_r t}$ appears with again $n=2$ \cite{Mayer2021,Han2006}.

\section{Brownian dynamics simulation}

\begin{figure*}[ht]
\includegraphics[scale=0.8]{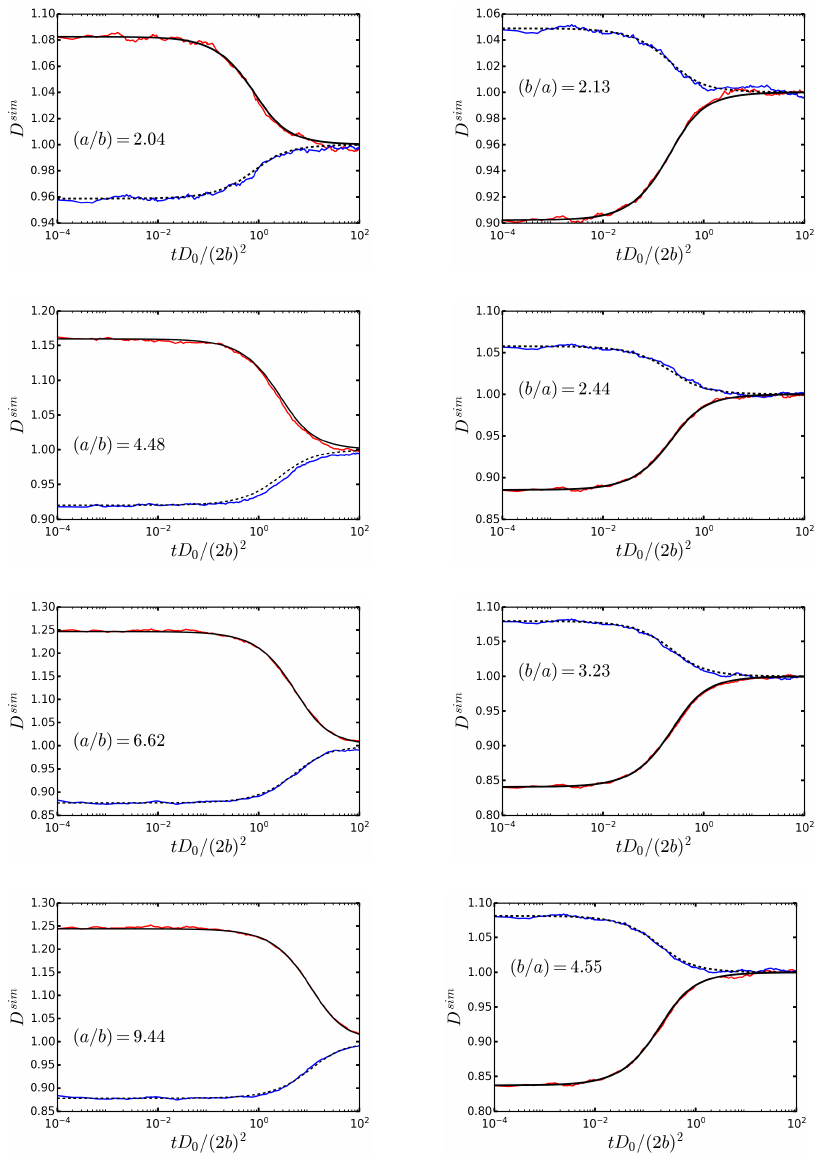}
\caption{Simulations: The parallel and perpendicular translation diffusion coefficients obtained from Brownian dynamics simulations for prolate and oblate ellipsoids at dilute densities. For prolate (oblate) case, the curve of $D^{sim}_{||}$ is higher (lower) than the curve of $D^{sim}_{\perp}$ at short times.}
    \label{diffusionInSimulations}
\end{figure*}

In order to test the theory and better understand the experimental results, we carried out Brownian dynamics simulations using LAMMPS \cite{LAMMPS}. The idea here is to use the experimental diffusion tensors, both for translations and rotations, as inputs for the Brownian simulation algorithm in LAMMPS \cite{Ilie2015, Delong2015} for ellipsoids at different aspect ratios (prolate and oblate systems). 
 
From Figs.~\ref{diffusionInSimulations}, we can see that the short-time parallel and perpendicular translation diffusions in simulations agree with the theory for all the aspect ratios. The plots of the simulations confirm the fact that the orientation memory is lost with time. The parallel and perpendicular translation diffusions are not identical at short times (anisotropic diffusion), but they become identical at large times (isotropic diffusion). This originates in the translation-rotation coupling in Perrin's equations. 
 For all aspect ratios tested, we may conclude that the experimental data have been taken at such strong dilution as to correspond to the single particle limit.
 
\end{document}